\documentclass[aps,nofootinbib,prl,reprint,superscriptaddress]{revtex4-1}
\usepackage{blindtext}
\usepackage{xcolor}
\usepackage{hyperref}
\usepackage{epsfig}
\usepackage{subfigure}
\usepackage{graphicx}
\usepackage[toc,page]{appendix}
\usepackage{epstopdf}
\usepackage{amsmath,amssymb,amsfonts}
\usepackage{array}
\usepackage{url}
\usepackage{hyperref}
\usepackage{lineno}
\usepackage{xspace}
\usepackage{listings}
\usepackage{float}
\usepackage{subfigure}

\newcommand{\bc}{\begin{center}}
\newcommand{\ec}{\end{center}}
\newcommand{\be}{\begin{equation}}
\newcommand{\ee}{\end{equation}}
\newcommand{\bea}{\begin{eqnarray}}
\newcommand{\eea}{\end{eqnarray}}
\newcommand{\bef}{\begin{figure}}
\newcommand{\ef}[1]{\label{#1}\end{figure}}
\newcommand{\bet}{\begin{table}}
\newcommand{\et}[1]{\label{tab#1}\end{table}}

\newcommand{\mub}{\mu_{\scriptscriptstyle B}}

\newcommand{\snn}{\sqrt{s_{_\mathrm{NN}}}}
\newcommand{\pt}{\ensuremath{p_{\rm T}}\xspace}
\newcommand{\ie}{{\sl i.e.\/}}
\newcommand{\RNum}[1]{\uppercase\expandafter{\romannumeral #1\relax}}

\definecolor{theblue}{RGB}{0,50,230}
\hypersetup{
  colorlinks=true,
  linkcolor=theblue,
  citecolor=theblue,
  urlcolor=theblue
}

\begin{document}

\title{Collision-energy Dependence of Deuteron Cumulants and Proton-deuteron Correlations in Au+Au collisions at RHIC}
\include{star-author-list-2024-03-28.listB_aps}

\begin{abstract}
We report the first measurements of cumulants, up to $4^{th}$ order, of deuteron number distributions and proton-deuteron correlations in Au+Au collisions recorded by the STAR experiment in phase-I of Beam Energy Scan (BES) program at the Relativistic Heavy Ion Collider. Deuteron cumulants, their ratios, and proton-deuteron mixed cumulants are presented for different collision centralities covering a range of center-of-mass energy per nucleon pair $\snn$~=~7.7 to 200~GeV. It is found that the cumulant ratios at lower collision energies favor a canonical ensemble over a grand canonical ensemble in thermal models. An anti-correlation between proton and deuteron multiplicity is observed across all collision energies and centralities, consistent with the expectation from global baryon number conservation. The UrQMD model coupled with a phase-space coalescence mechanism qualitatively reproduces the collision-energy dependence of cumulant ratios and proton-deuteron correlations.
\end{abstract}

\maketitle
\section{Introduction}
One of the major goals of heavy-ion collision experiments is to study the phases of strongly interacting nuclear matter versus temperature and pressure. Experimental results have demonstrated the existence of a deconfined state of quarks and gluons~\cite{BRAHMS:2004adc, PHOBOS:2004zne, STAR:2005gfr, PHENIX:2004vcz, CMS:2012bms, ALICE:2022wpn}. The mean yields of hadrons produced in central heavy-ion collisions can be described by thermal models with a suitable choice of chemical freeze-out parameters such as temperature ($T$) and baryon chemical potential ($\mub$). The typical values of $T$ vary from around 140~MeV at collision energy ($\snn$) of 7.7~GeV to 160~MeV at the energy of 5.02~TeV~\cite{STAR:2017sal, STAR:2003jwm, Andronic:2017pug}. However, deuterons, tritons, and other light nuclei, which have binding energies of the order of a few MeV, are also produced in heavy-ion collisions~\cite{STAR:2019sjh, ALICE:2015wav, NA49:2016qvu}. Interestingly, the yields of light nuclei can also be explained with temperatures similar to those extracted using hadronic yields~\cite{Andronic:2017pug, Chatterjee:2014ysa, Biswas:2020kpu}. The natural question that arises then is: how are light nuclei produced in a medium that freezes out at the temperature of the order of 100~MeV?

The production mechanism of light nuclei is commonly studied in two approaches: the {\it thermal model} and the {\it coalescence model}. The thermal model treats light nuclei as any other hadrons and their masses and quantum numbers are inputs to the model. These model calculations show good agreement with experimental data on transverse momentum ($\pt$) integrated mid-rapidity yields of deuterons and deuteron to proton yield ratios in central heavy-ion collisions~\cite{Andronic:2017pug, STAR:2019sjh}. In the coalescence model, light nuclei are formed by coalescing protons and neutrons with a finite probability determined by their closeness to each other in the phase-space~\cite{Butler:1963pp, Kapusta:1980zz}. One of the signatures of the coalescence mechanism is that the elliptic flow of light nuclei should show constituent nucleon number scaling~\cite{Yan:2006bx}, and such a scaling property has been observed in the STAR experiment~\cite{STAR:2016ydv}. Both the thermal and coalescence models have been fairly successful in explaining the set of experimental data. However, the production mechanism of light nuclei still needs to be understood in detail~\cite{Mrowczynski:1987oid, Scheibl:1998tk, Andronic:2010qu, Mrowczynski:2016xqm, Chatterjee:2014ysa}. It is not necessarily true that deuteron production has to happen only via one of the above-mentioned mechanisms. Both mechanisms might be at work in heavy-ion collisions~\cite{Kapusta:1980zz}.

Furthermore, higher order cumulants of particle multiplicity distributions are known to probe finer details of the thermodynamics of the system created~\cite{Stephanov:1999zu, Karsch:2010ck, Borsanyi:2014ewa, Alba:2014eba, Gupta:2022phu, Pandav:2022xxx}. Recent studies suggest that cumulants of event-by-event deuteron number distribution might have different signatures in thermal and coalescence approaches and can shed light on their production mechanism~\cite{Feckova:2016kjx}. Calculations using a simple coalescence model predict the rise of cumulant ratios towards lower collision energies in contrast to the predictions from the thermal model using Grand Canonical Ensemble (GCE) and the Poisson baseline, both of which are equal to 1 across collision energies~\cite{Feckova:2016kjx}.

In addition to probing the production mechanism, higher moments of deuteron number fluctuation can potentially be sensitive to signals of the QCD critical point, and first-order phase transition. Even though deuteron has a binding energy of only 2.2~MeV, its production is predicted to be affected by the enhancement of pre-clustering of nucleons at the chemical freeze-out due to modifications in the nucleon-nucleon interaction near a phase transition~\cite{Shuryak:2018lgd, Shuryak:2019ikv}. Also, a certain combination of the proton, deuteron, and triton yields is constructed to probe neutron density fluctuations at the kinetic freeze-out~\cite{Sun:2017xrx} and has been measured by the STAR experiment. These results show an excess over the coalescence baseline in central collisions at $\snn$~=~19.6 and 27~GeV~\cite{STAR:2022hbp}. Further, as deuterons carry two baryons, their fluctuation may add to the understanding of the baryon number fluctuations in heavy-ion collisions.\\

In this paper, we report the first measurements of the cumulants of the deuteron multiplicity distribution and the proton-deuteron number correlation from Au+Au collisions recorded by the STAR detector~\cite{STAR:2002eio} at Relativistic Heavy Ion Collider (RHIC) from the years 2010 to 2017. The data are presented for Au+Au collisions at $\snn$~=~7.7, 11.5, 14.5, 19.6, 27, 39, 54.4, 62.4, and 200~GeV corresponding to a wide range of baryon chemical potential from 420 to 20~MeV~\cite{STAR:2017sal, Cleymans:2005xv}. These results are compared to several model calculations: a thermal model using Grand Canonical and Canonical Ensembles (GCE and CE)~\cite{Vovchenko:2019pjl}, the Ultrarelativistic Quantum Molecular Dynamics (UrQMD) model~\cite{Bleicher:1999xi, Petersen:2008kb} combined with a phase-space coalescence mechanism~\cite{Sombun:2018yqh}, and a simple coalescence model from Ref~\cite{Feckova:2016kjx}.

\section{\label{sec:obs}Observables}
Distributions can be characterized by their cumulants of various order. A general expression to find any order cumulants of a distribution can be found in ~\cite{STAR:2021iop}. The cumulants ($C_{n}$) up to order $n$~=~4 are defined below. We use $N$ to represent the number of deuterons in one event and $\langle N \rangle$ for the average value over the entire event ensemble. Then the deviation of $N$ from its event average is given by $\delta N = N - \langle N \rangle$.
\begin{eqnarray}
C_1 &=& \langle N \rangle\\
C_2 &=& \langle(\delta N)^2\rangle\\
C_3 &=& \langle(\delta N)^3\rangle\\
C_4 &=& \langle(\delta N)^4\rangle -3 \langle(\delta N)^2\rangle^2
\label{eqn_cumu}
\end{eqnarray}
The moments can be expressed in terms of the cumulants:
\begin{equation}
M = C_1 \ ,\ \ \sigma^{2} = C_2 \ , \ \  S = \frac{C_3}{C_{2}^{3/2}}\ , \ \ \kappa = \frac{C_{4}}{C_{2}^{2}} \ ,
\label{eqn_moments}
\end{equation}
where $M$ is the mean, $\sigma$ is the standard deviation, $S$ is the skewness and $\kappa$ is the kurtosis.

To eliminate the system volume dependence of cumulants, their ratios are usually constructed as follows~\cite{Gupta:2009mu}:
\begin{equation}
\frac{\sigma^{2}}{M} = \frac{C_{2}}{C_{1}} \ , \ \ S\sigma = \frac{C_{3}}{C_{2}}\ , \ \ \kappa \sigma^{2} = \frac{C_{4}}{C_{2}} \ .
\label{eqn_cumuratio}
\end{equation}
These ratios can be connected to the ratios of number susceptibilities calculated in thermal models~\cite{Karsch:2010ck} as $C_{2}/C_{1} = \chi_{2}/\chi_{1}$, $C_{3}/C_{2} = \chi_{3}/\chi_{2}$, and $C_{4}/C_{2} = \chi_{4}/\chi_{2}$. The $n$-th order number susceptibility is $\chi_{n} = d^{n}[P/T^{4}]/d(\mu/T)^{n}$, where $P$, $T$, and $\mu$ are the pressure, temperature, and chemical potential, respectively.

If the particle multiplicity follows the Poisson distribution, cumulants of all orders are equal and therefore their ratios are unity. Poisson expectations are used as the statistical baselines for the measured cumulant ratios.

The Pearson correlation coefficient measures the linear correlation between two variables. The correlation coefficient between proton and deuteron numbers can be defined as follows:
\begin{equation}
\frac{C^{(1,1)}_{(p,d)}}{\sigma_{p}\sigma_{d}} = \frac{\langle(\delta N_{p}\delta N_{d})\rangle}{\sigma_{p}\sigma_{d}} = \frac{\langle N_{p} N_{d}\rangle - \langle N_{p}\rangle \langle N_{d}\rangle}{\sigma_{p}\sigma_{d}} \ ,
\label{eqn_pearson}
\end{equation}
where $N_{p}$ and $N_{d}$ are proton and deuteron numbers, respectively. The correlation coefficient ranges from $-1$ to $1$. A positive sign of the coefficient implies that two variables are correlated while a negative sign implies an anti-correlation. A zero value of the coefficient implies that two variables are uncorrelated.

\section{\label{sec:anam}Analysis methods}
\begin{figure*}[hbt]
\bc\includegraphics[scale=0.55]{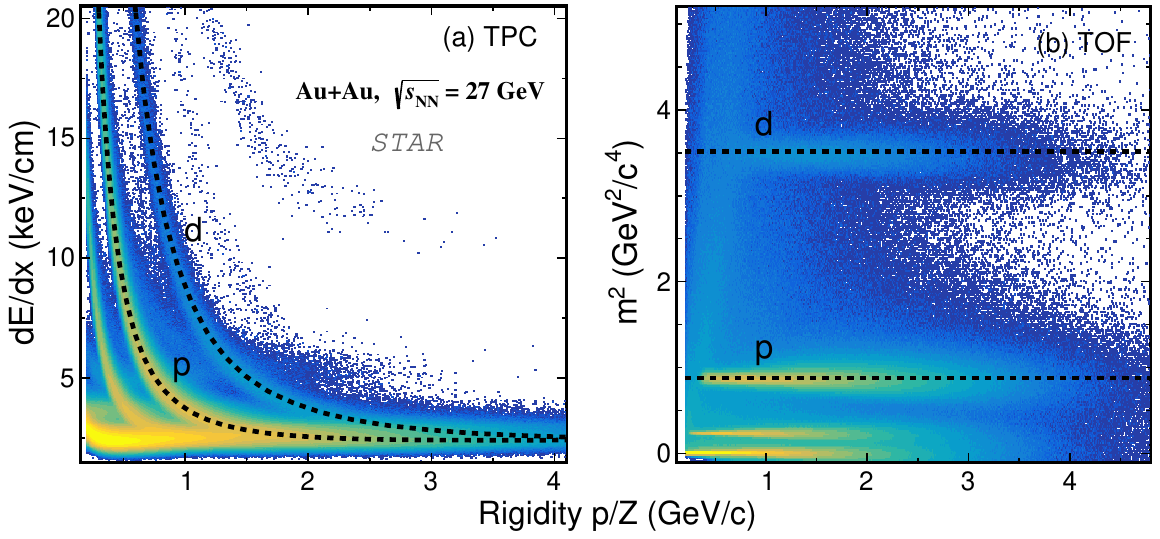}\ec
\caption{$\left\langle dE/dx \right\rangle$ and $m^{2}$ distribution of charged particles for $|\eta| \ <$~1.0 in Au+Au collisions at $\snn$~=~27~GeV. Panel (a): The $\left\langle dE/dx \right\rangle$ distribution of charged particles from TPC~\cite{Anderson:2003ur} as a function of rigidity ($p/Z$). The dashed curves represent the expected values of $\left\langle dE/dx \right\rangle$ calculated using the Bichsel function~\cite{Bichsel:2006cs} for the corresponding particles. Panel (b): Mass squared of charged particles as a function of momentum from TOF~\cite{Llope:2005yw}. The dashed lines represent the mass squared values for the corresponding particles.}\label{dedx_m2}
\end{figure*}

The results presented here are measured in minimum-bias~\cite{STAR:2017sal} Au+Au collisions at $\snn$~=~7.7 to 200~GeV recorded using the STAR detector at RHIC. Collision events are selected having the vertex position ($V_{z}$) within $\pm$30~cm ($\pm$40~cm for $\snn$~=~7.7~GeV) with respect to the nominal center of the STAR Time Projection Chamber (TPC) detector along the beam direction ($z$ axis). Events at each collision energy are further divided into centrality classes using the produced charged particle multiplicity as a measure. Central collision events have higher values of charged particle multiplicity compared to peripheral collision events. The charged particle multiplicity used for the centrality classification is selected using the TPC detector with pseudorapidity ($\eta$) within $-1$ to $+1$. Protons, deuterons, and their anti-particles are removed from the definition of collision centrality. This avoids the self-correlation effect between deuterons used to calculate cumulants and particles included in the centrality definition~\cite{Luo:2013bmi, Chatterjee:2019fey, STAR:2020tga, STAR:2021iop}. The results presented here correspond to three event classes: most central collisions (events from the top 5\% of the above-mentioned multiplicity distribution), mid-central (events from 30-40\% of the distribution), and peripheral collisions (events from 70-80\% of the distribution).
The number of analyzed events for minimum bias collisions at each energy is provided in Table~\ref{tab1_stats}.
\begin{table}[!htb]
\small
\caption{\small Total number of minimum bias events for Au+Au collisions analyzed for various collision energies obtained after all the event selection cuts are applied.}
	\centering   
 \small
	\resizebox{\columnwidth}{!}{
 \begin{tabular}{|c|c|c|c|c|c|c|c|c|c|}
		\hline
		$\sqrt{s_{NN}}$ (GeV) &  7.7 & 11.5 & 14.5 & 19.6 & 27 & 39 & 54.4 & 62.4 & 200 \\
		\hline
		Events (Millions) &  2.2 & 6.6 & 12 & 14 & 30 & 83 & 520 & 37 & 220 \\
		\hline
	\end{tabular}
 }
	\label{tab1_stats}
\end{table}
The charged tracks used for the cumulant analysis are required to have more than 20 space points in the TPC to ensure good track momentum resolution and the ratio between assigned to total possible space points is taken to be greater than 0.52 in order to minimize track splitting. The distance of the closest approach (DCA) of the selected tracks to the primary vertex is required to be within 1~cm in order to suppress contamination from secondary particles~\cite{STAR:2007zea, STAR:2008med}. To identify deuterons and protons, particle identification (PID) selection criteria are further applied to the charged tracks. PID is done via ionization energy loss ($dE/dx$) measured in the TPC~\cite{Anderson:2003ur} and mass squared ($m^{2}$) obtained from the Time Of Flight (TOF)~\cite{Llope:2005yw} detectors. Panel (a) in Fig.~\ref{dedx_m2} shows the measured $\langle dE/dx \rangle$ vs. rigidity (\ie ~momentum/charge) of particles in $|\eta|\ <$~1.0. Various bands corresponding to particles of different masses are clearly separated at low momentum. An extension of PID to higher $\pt$ is achieved by using the TOF detector. Panel (b) in Fig.~\ref{dedx_m2} shows the distribution of $m^{2}$ calculated using the information (path length and time of travel by the particle) from the TOF detector. The kinematic region for deuterons covers the full azimuth range, mid-rapidity ($|y|\ <$~0.5), and the $\pt$ range is from 0.8 to 4~GeV/$c$. Both TPC and TOF are used to get good purity, above 98\%, of the deuteron sample. For proton-deuteron correlation measurement, protons are identified at mid-rapidity with $\pt$ between 0.4 and 2.0~GeV/$c$. To ensure good efficiency for the proton sample, for the $\pt$ range $0.4<\pt<0.8$~GeV/$c$, only TPC is used while both TPC and TOF detectors are simultaneously used for the range $0.8<\pt<2.0$~GeV/$c$~\cite{STAR:2021iop}. For the momentum ranges studied, the typical value of the TPC tracking (TOF matching) efficiency for deuterons in 0-5\% most central collisions at $\snn$~=~7.7~GeV is 81\% (69\%). The corresponding values at $\snn$~=~200~GeV are 63\% (64\%). Protons are identified with similar values of detection efficiencies~\cite{STAR:2021iop}.

The cumulants are corrected for finite track reconstruction efficiency in the TPC and track matching efficiency in TOF detectors. The correction is performed assuming a binomial response of both detectors for deuteron and proton efficiencies~\cite{Nonaka:2017kko}. In addition, cumulants are corrected for their dependence on multiplicity by using the Centrality Bin-Width Correction (CBWC) method~\cite{Luo:2013bmi} for each centrality. This correction suppresses the effect of initial volume fluctuations on the measured cumulants arising due to fluctuations in the impact parameter of collisions.

The statistical uncertainties on the measurements are calculated using a Monte Carlo approach called the Bootstrap method~\cite{Luo:2014rea, Pandav:2018bdx}. The systematic uncertainties are estimated by varying the track selection and particle identification criteria. Track quality cuts such as DCA, the number of space points in the TPC, and PID criteria such as cuts on measured $dE/dx$ and $m^{2}$ values are considered as the sources of systematic uncertainty~\cite{STAR:2021iop}. In addition, a $\pm 5$\% uncertainty associated with the reconstruction efficiency of the detector is also included in the overall systematic uncertainty. For each source of systematics, the standard deviation from the default set of results is calculated. The systematic uncertainty is determined from the square root of the quadratic sum of the standard deviations coming from different sources. The typical systematic errors, for example in 0-5\% most central collisions at 7.7~GeV, are of the order of 5\% for $C_{1}$, $C_{2}$, and $C_{3}$ and 6\% for $C_{4}$. The uncertainty in the reconstruction efficiency estimation makes the biggest contribution to the systematics.

\section{\label{sec:res}Results and discussion}
\bef[tbh!]
\vspace{0.2cm}
\bc\includegraphics[scale=0.46]{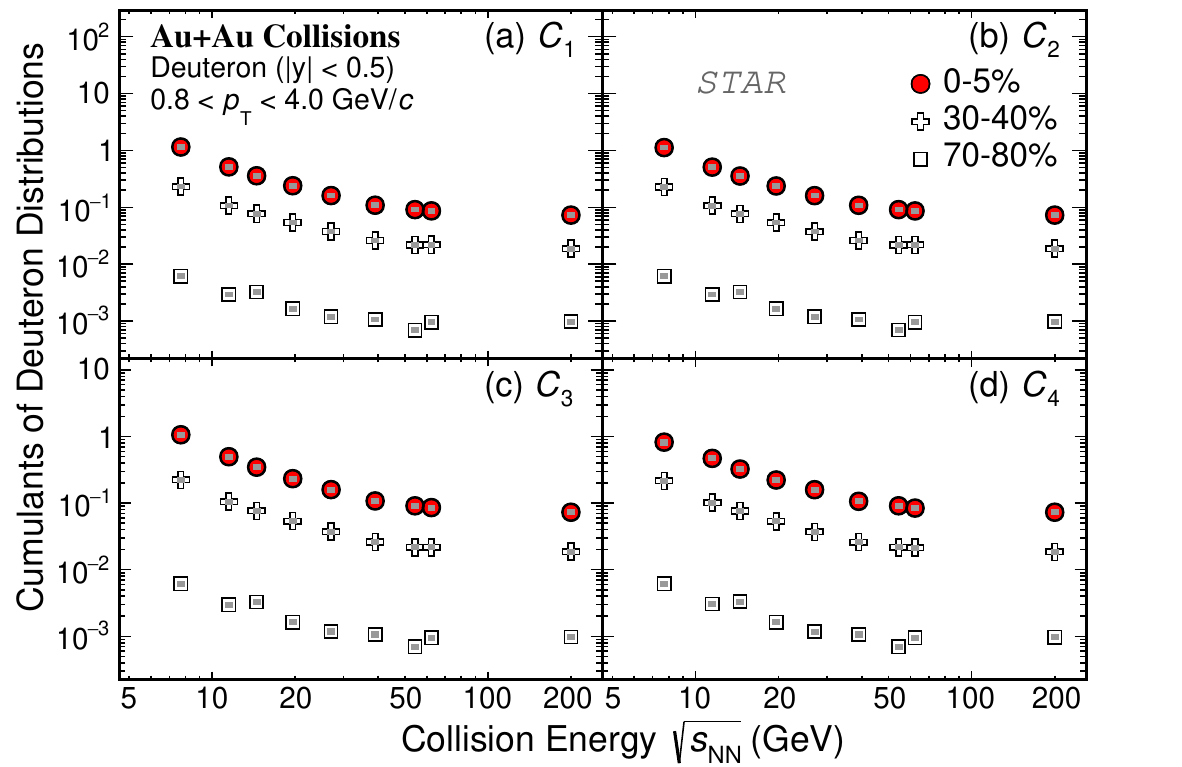}\ec 
\caption{Cumulants ($C_{n}, \  n=1-4$) of the deuteron distributions as a function of collision energy for most central, mid-central, and peripheral Au+Au collisions as measured by STAR.
Cumulants are corrected for finite detector efficiencies~\cite{Nonaka:2017kko} and centrality bin-width effect~\cite{Luo:2013bmi}. Uncertainties on the cumulants are smaller than marker symbols. Results for most central, mid-central, and peripheral collisions are shown using solid circle, open cross, and open square markers, respectively. Bar and cap symbols represent the statistical and systematic uncertainties, respectively. The transverse momentum range for the measurements is from 0.8 to 4~GeV/$c$ and the rapidity range is $-0.5<y<0.5.$}\label{cumu_plot}
\ef{cumu}
Figure~\ref{cumu_plot} shows the deuteron cumulants ($C_{n}, \ n=1-4$) as a function of $\snn$ for most central (0-5\%), mid-central (30-40\%), and peripheral (70-80\%) Au+Au collisions. The cumulants $C_{1}$ to $C_{4}$ of deuteron distributions for most central Au+Au collisions smoothly increase with decreasing $\snn$. This indicates an enhanced production of deuterons towards the high baryon density region (corresponding to low $\snn$~\cite{STAR:2017sal}). The effect of high baryon density on deuteron production can be understood using a thermal model. In the thermal model, baryon density dependence is given by the factor $\sim \exp[(B\mub - m_{d})/T]$, where $B$ and $m_{d}$ are the baryon number and mass of the deuteron, respectively. As light nuclei carry multiple baryons, the contribution of the above factor is especially enhanced in the high baryon density region~\cite{Cleymans:2011pe}. Cumulants in the mid-central and peripheral collisions show a similar $\snn$ dependence as seen for the most central collisions. For any given $\snn$, the cumulants of any order increase from peripheral to central collisions. For $\snn$~=~27~GeV and above, in any given collision energy and centrality, $C_{1}$ to $C_{4}$ values are close to each other and almost independent of order ($n$) of the cumulant. This implies that the event-by-event deuteron number distribution at higher $\snn$ exhibit a near-Poissonian behavior.
\bef[tbh!]
\vspace{0.2cm}
\bc\includegraphics[scale=0.46]{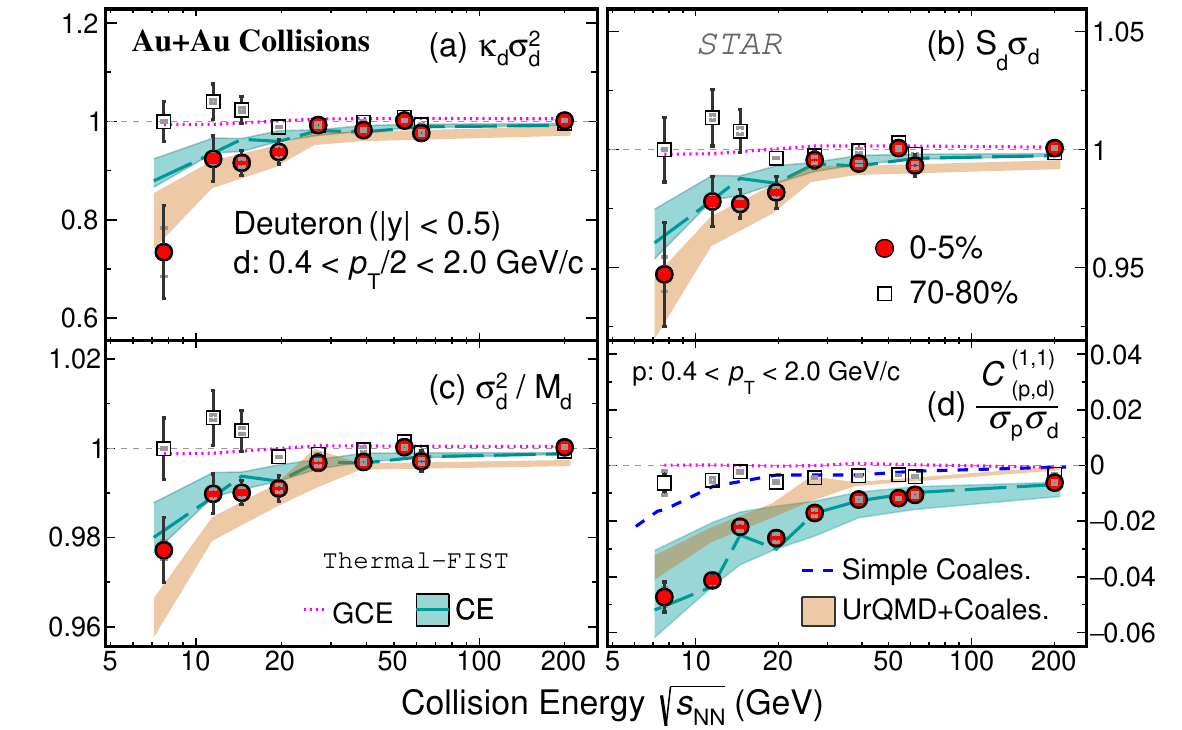}\ec
\caption{Cumulant ratios of deuteron distributions and proton-deuteron correlation shown as a function of collision energy. Red circle and open square markers represent measurements for most central (0-5\%) and peripheral (70-80\%) collisions, respectively. Bar and cap symbols represent the statistical and systematic uncertainties, respectively. The gray dashed line is the Poisson baseline (unity for cumulant ratios and zero for correlation). All model results presented in the figure correspond to the most central (0-5\%) collisions. Calculations from an UrQMD coupled with a phase-space coalescence model~\cite{Sombun:2018yqh} are shown using the orange color-filled band (the width of the band represents the statistical uncertainty). Thermal-FIST~\cite{Vovchenko:2019pjl} calculations for GCE are shown using a magenta dashed line. The cyan color-filled band represents the CE thermal model results corresponding to the range of canonical correlation volume ($V_{c}$) from $2dV/dy$ to $4dV/dy$. CE thermal model results for $\chi^{2}$ minimum fit of the above-mentioned four observables is shown using a cyan color dashed line. In panel (d), predictions for one of the assumptions in a simple coalescence model from Ref.~\cite{Feckova:2016kjx} are shown using a blue dashed line.}\label{cumuratio_plot}
\ef{cumuratio}
Figure~\ref{cumuratio_plot} shows the collision energy dependence of the cumulant ratios and the proton-deuteron number Pearson correlation coefficient for most central 0-5\% and peripheral 70-80\% Au+Au collisions. The cumulant ratios $\kappa\sigma^{2}$, $S\sigma$, and $\sigma^{2}/M$ in central collisions show smooth dependence on collision energy. At higher colliding energies ($\snn \geq $~27~GeV), most central 0-5\% cumulant ratios are close to the Poisson baseline (unity) and deviate from unity as $\snn$ decreases. In low-energy collisions, cumulants are increasingly suppressed with increasing order $n$, resulting in the $\kappa\sigma^{2}$ showing the largest deviation from unity compared to the other two ratios which involve lower-order cumulants. Note that the scales of the y-axis are different in different panels. The observed suppression of cumulant ratios might arise because of global baryon number conservation, which can notably affect the measurements performed at mid-rapidity in low-energy collisions. In low-energy collisions ($\snn < $~27~GeV), due to an increase in the number of net baryons at mid-rapidity~\cite{Videbaek:1995mf} and the acceptance cuts which include a larger fraction of the phase space, one observes an enhanced effect of baryon number conservation~\cite{Braun-Munzinger:2020jbk}. As the fraction of net baryons in the measurement acceptance over the total net-baryon numbers produced in the collision increases, the event-by-event fluctuations of deuterons become constrained. A model calculation with the canonical effect implemented via local conservation of baryon number is shown to have a small impact on higher order net-proton cumulants~\cite{Braun-Munzinger:2019yxj}. However, model studies with global baryon number conservation show that the suppression increases with the order of the cumulants and also increases with decreasing collision energies, as observed in our measurements~\cite{Braun-Munzinger:2020jbk}. Corresponding results in 70-80\% peripheral centrality show a weak dependence on collision energy and are close to unity. Cumulant ratios for peripheral collisions are found to be least affected by the global baryon number conservation. Cumulant ratio values in 30-40\% centrality lie between those for most central and peripheral collisions~\footnote{Data points for 30-40\% centrality are not shown in Fig.~\ref{cumuratio_plot} to avoid clutter. Relevant results can be found in the HEPData.}.

Calculations from the Thermal-FIST~\cite{Vovchenko:2019pjl} model for the most central 0-5\% collisions are also shown in Fig.~\ref{cumuratio_plot}. This model assumes an ideal gas of hadrons, resonances, and light nuclei in thermodynamic equilibrium. Excited states of light nuclei which decay to protons and deuterons could also be included in the particle list for the thermal model. However, as seen in Ref.~\cite{Vovchenko:2020dmv}, the contribution from excited nuclei feed down is very small for $\snn \geq $ 7.7~GeV, and is not taken into account in our model calculation. Thermal model calculations are performed for both grand canonical and canonical ensembles and the experimental acceptances have been taken into account. The input parameters of the model such as chemical freeze-out temperature, chemical potentials, and kinetic freeze-out conditions are taken from Ref.~\cite{STAR:2017sal} which are extracted from thermal model fits of hadronic yields and $\pt$-spectra measured in the STAR experiment. Results for the cumulant ratios from the GCE framework of the Thermal-FIST model are close to unity across all collision energies. 

The GCE model seems to fail to describe the ratios for $\snn \leq $ 20~GeV. The CE thermal model which incorporates baryon number conservation, predicts the suppression of cumulant ratios as observed in the data. Note that in the CE model, only the canonical effect due to the conservation of baryon number is considered for light nuclei fluctuations. The canonical ensemble in the Thermal-FIST model uses an additional volume parameter called the canonical correlation volume, $V_{c}$, over which the exact conservation of the baryon number is implemented. The shaded band represents the results for $V_{c}$ in the range of 2 to 4 times the $dV/dy$, where $dV/dy$ is the chemical freeze-out volume per unit rapidity that is obtained from the thermal model fit of hadronic yields~\cite{STAR:2017sal}. The model parameter $V_{c}$ is also varied at each collision energy for a reasonable agreement with the measured values of $\kappa\sigma^{2}$, $S\sigma$, $\sigma^{2}/M$, and the Pearson coefficient. The line shows the results corresponding to minimum $\chi^{2}$ fits by scanning the $V_{c}$ parameter in the model. $V_{c}$ values are found to vary from $2dV/dy$ at the lowest energy to $4dV/dy$ at the highest RHIC collision energy. A slightly higher range of $V_{c}$ is obtained at LHC energies for measurements from the ALICE collaboration~\cite{Vovchenko:2019kes, Vovchenko:2020kwg}. The higher value of canonical correlation volume implies that the part of the system under measurement is approaching the grand-canonical limit~\cite{Vovchenko:2019kes}. This also highlights the importance of the canonical ensemble thermal model at lower collision energies.  

Physics mechanisms such as decay of resonances~\cite{Bluhm:2016byc} and transport of beam protons to mid-rapidity~\cite{Bzdak:2016jxo} also could affect the cumulants. For this, we compare results from the UrQMD model (v3.4 in default cascade mode) combined with a phase-space coalescence mechanism to the experimental data. The UrQMD model is a hadronic transport code which takes into account many physics mechanisms including those from transport of beam protons to mid-rapidity, resonance decays, binary scattering of hadrons, string dynamics, and conservation of net-baryon number~\cite{Bass:1998ca}. Phase space information of protons and neutrons at the kinetic freeze-out surface from the UrQMD model are used as inputs to the coalescence mechanism to form deuterons. In the coalescence model, proton-neutron pairs with relative momenta within 0.285~GeV/$c$ and position space separations within 3.575~fm are considered as candidates for deuteron formation. These parameters in model studies~\cite{Sombun:2018yqh} are found to provide a good description of experimental data on deuteron yields. The UrQMD model combined with the coalescence mechanism, also reproduce the energy dependence trend observed in data and show a fair agreement with the measured cumulant ratios.

In panel (d) of Fig.~\ref{cumuratio_plot}, we observe that the Pearson correlation coefficient between proton and deuteron numbers is negative across all collision energies and centralities presented, which implies that the proton and deuteron numbers are anti-correlated with each other. At lower colliding energy, anti-correlation becomes stronger for most central 0-5\% Au+Au collisions. These measurements for peripheral Au+Au collisions do not show any energy dependence and are close to the statistical expectations. In the GCE thermal model, protons and deuterons are uncorrelated. However, the CE thermal model calculation correctly predicts the sign and energy dependence trend of the measured correlation. Predictions from a simple coalescence model from Ref.~\cite{Feckova:2016kjx} are also shown for the most central Au+Au collisions. For simplicity, the authors of Ref.~\cite{Feckova:2016kjx} assume Poisson distributions for both protons and neutrons, with their numbers fluctuating independently. Note that this model does not take into account the details of the phase space information of coalescing protons and neutrons. On the other hand, the fair agreement of predictions from the UrQMD model combined with the phase-space coalescence mechanism~\cite{Sombun:2018yqh} with the experimental data in most central 0-5\% collisions suggests that the phase space information of constituent nucleons is important for the deuteron formation process in the coalescence mechanism. The ALICE collaboration recently reported measurements on proton-deuteron correlation in Pb+Pb collisions at $\snn$~=~5.02~TeV. The Pearson correlation coefficient was found to have small negative values and is mostly constant for all collision centralities~\cite{ALICE:2022xiu}. Similar to the observations of this study, the CE thermal model calculations with baryon number conservation implemented also explain the ALICE data for suitable choices of model parameters. The negative sign of the Pearson correlation coefficient across the range of collision energies (GeV to TeV) and centralities (central to peripheral) establishes the importance of baryon number conservation in baryon-nucleus correlations. The nature of the agreement of the proton-deuteron correlation data with the CE thermal model calculation suggests a canonical thermal effect over a coalescence mechanism. At the same time, there is reasonable scope for improvements in both the production models discussed here.

\bef[tb!]
\bc\includegraphics[scale=0.38]{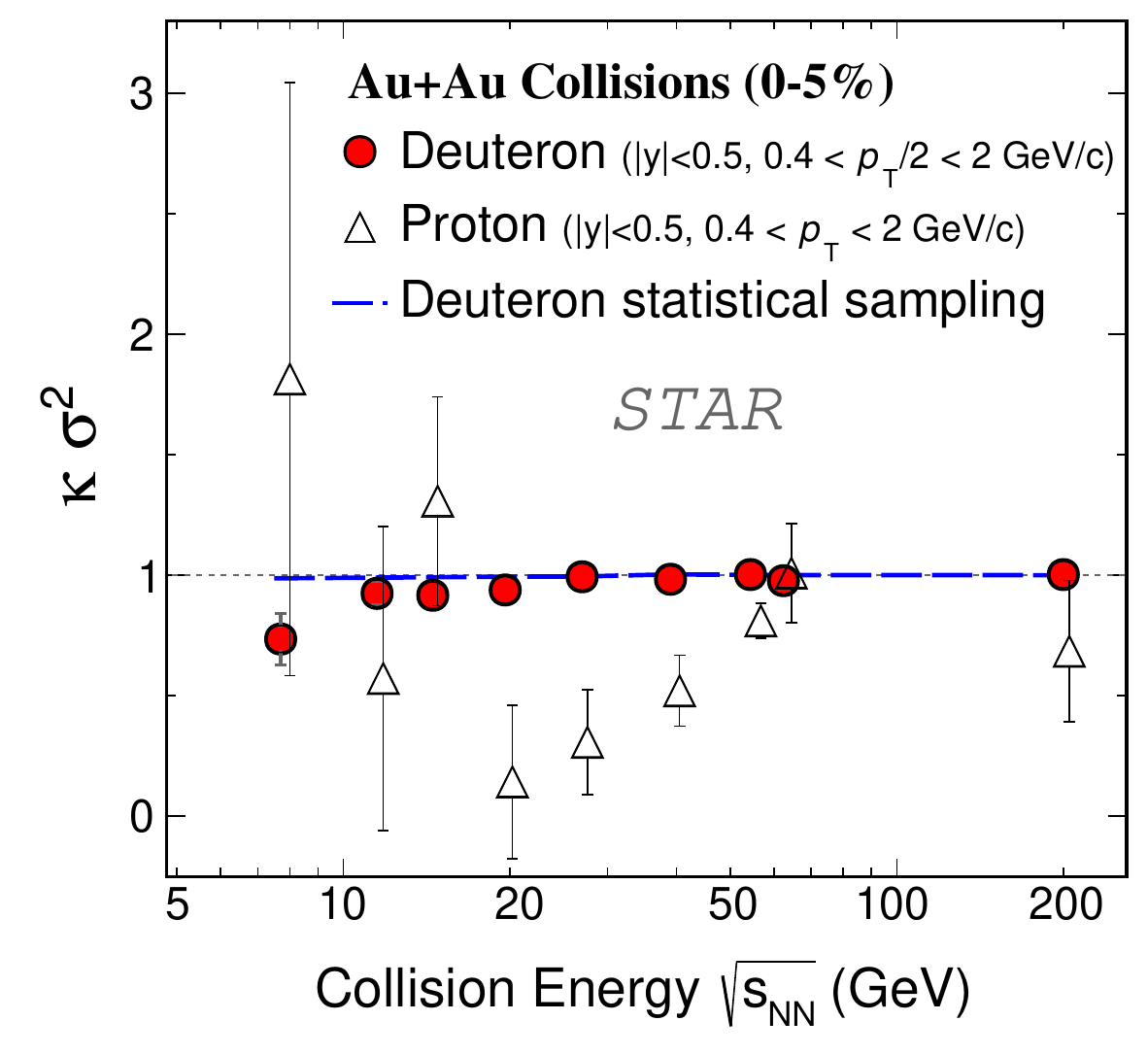}\ec 
\caption{$\kappa \sigma^{2}$ of deuteron
and proton distribution for most central (0-5\%) Au+Au collisions. Red circle and black triangle markers represent deuteron and proton data~\cite{STAR:2021iop}, respectively. The gray dashed line is the Poisson baseline (unity). $\kappa \sigma^{2}$ of deuterons show a smooth dependence on the collision energy in contrast to
protons.}\label{kurtosis_plot}
\ef{kurtosis}

As deuterons carry two baryons, it is important and interesting to investigate how their cumulant ratios differ from those of the protons. Figure~\ref{kurtosis_plot} shows the comparison of $\kappa\sigma^{2}$ of the deuteron multiplicity distribution to those of protons~\cite{STAR:2021iop} for most central 0-5\% Au+Au collisions. For the $\kappa\sigma^{2}$ of protons, the larger statistical uncertainties are attributed to the larger width of proton multiplicity distributions as compared to the deuteron distributions~\cite{Luo:2014rea}. Within the current uncertainties, the proton $\kappa\sigma^{2}$ (similar to that of net-proton) shows a non-monotonic $\snn$ dependence~\cite{STAR:2020tga} in most central Au+Au collisions. This feature is similar, at a qualitative level, to the theoretical predictions near the QCD critical point. The $\kappa\sigma^{2}$ for deuterons, however, shows a weaker dependence on collision energy compared to that for protons. This could be due to deuterons having a very low event-by-event yield compared to protons, resulting in reduced sensitivity to any possible critical point physics. To test the effect of low event-by-event yield on the cumulant ratios, a simple statistical simulation is utilized by using the measured deuteron to proton yield ratios~\cite{STAR:2019sjh} and proton cumulants~\cite{STAR:2021iop} as inputs. Using a two-component function, which is a superposition of Poisson and binomial distributions (originally developed in Ref.~\cite{Bzdak:2018uhv} for a different purpose), the proton distribution is modeled in order to reproduce the measured proton cumulants in most central 0-5\% Au+Au collisions. Then deuteron multiplicity on an event-by-event basis is sampled from the above-mentioned proton distribution using the $d/p$ ratio~\cite{STAR:2019sjh} as the binomial probability of success to form a deuteron. The $\kappa\sigma^{2}$ calculated from this resultant deuteron distribution (shown in Fig.~\ref{kurtosis_plot} as a blue dashed line) is near unity and close to the experimental data. This test shows that the low deuteron multiplicity likely is responsible for the deuteron $\kappa\sigma^2$ being close to 1.
\section{\label{sec:sum}Summary}
We have presented measurements of deuteron cumulants, their ratios, and proton-deuteron number correlation performed in Au+Au collisions with the STAR detector at RHIC, covering a wide range of baryon chemical potential ($\mub$ from $\sim$ 20 to 420~MeV). The cumulant ratios of deuterons in most central collisions vary smoothly as a function of the collision energy and are suppressed below the Poisson baseline as the colliding energy decreases. The peripheral collision results, however, remain overall flat as a function of $\snn$. Anti-correlation between proton and deuteron numbers is observed across all collision energies and centralities studied. This anti-correlation becomes stronger for most central Au+Au collisions as the beam energy decreases. Cumulant ratios and correlations in mid-central collisions show a weaker dependence on collision energies compared to central collisions. These measurements for peripheral Au+Au collisions do not show a significant energy dependence and are close to the Poisson baseline. 

Important observations from the comparison of our measurements to the different model calculations can be summarized as follows. In most central Au+Au collisions, for thermal models: (i) GCE and CE reasonably describe the deuteron number fluctuation measurements above collision energies of 20~GeV. Only the CE model correctly predicts the negative sign of the proton-deuteron correlation. (ii) The thermal model with CE qualitatively agrees with the cumulant ratios for collision energies below 20~GeV, while the thermal model with GCE fails. As the CE model explicitly conserves the baryon number, this study reflects the importance of the role of conservation in fluctuation studies at lower collision energies. 

The UrQMD model coupled with a phase-space coalescence mechanism also describes the deuteron number fluctuation and deuteron-proton correlation measurements across all collision energies. A simple modeling of the coalescence process without taking into account the phase-space information of constituent nucleons fails to describe the measured proton-deuteron number correlation.

The $\kappa\sigma^{2}$ of the deuteron number distribution shows a smoothly decreasing trend with decreasing collision energy in contrast to protons. A simple statistical test suggests that the low deuteron multiplicity may be responsible for the observed near-Poisson behavior of deuteron cumulant ratios. Such trends as observed in the data currently do not support a scenario of enhanced formation of pre-clusters that might arise due to the presence of a CP/first-order phase transition. Our measurements can be utilized further to study the chemical freeze-out thermodynamics of deuterons and to constrain the light nuclei production model parameters. In the future, with higher event statistics and improved acceptance achieved in phase-II of BES and fixed-target collision datasets, $\pt$ and rapidity differential measurements with better statistical and systematic precision are possible. Further, fluctuations and hadron-nuclei correlation measurements can be performed for light nuclei species such as the triton, $^{3}\mathrm{He}$, and $^{4}\mathrm{He}$. This has the potential for a major improvement in the discriminating power of comparisons with model calculations and might help resolve the nuclei production puzzle in high-energy heavy-ion collisions.\newline

\noindent{\bf Acknowledgements:} We thank the RHIC Operations Group and RCF at BNL, the NERSC Center at LBNL, and the Open Science Grid consortium for providing resources and support.  This work was supported in part by the Office of Nuclear Physics within the U.S. DOE Office of Science, the U.S. National Science Foundation, National Natural Science Foundation of China, Chinese Academy of Science, the Ministry of Science and Technology of China and the Chinese Ministry of Education, the Higher Education Sprout Project by Ministry of Education at NCKU, the National Research Foundation of Korea, Czech Science Foundation and Ministry of Education, Youth and Sports of the Czech Republic, Hungarian National Research, Development and Innovation Office, New National Excellency Programme of the Hungarian Ministry of Human Capacities, Department of Atomic Energy and Department of Science and Technology of the Government of India, the National Science Centre and WUT ID-UB of Poland, the Ministry of Science, Education and Sports of the Republic of Croatia, German Bundesministerium f\"ur Bildung, Wissenschaft, Forschung and Technologie (BMBF), Helmholtz Association, Ministry of Education, Culture, Sports, Science, and Technology (MEXT), Japan Society for the Promotion of Science (JSPS) and Agencia Nacional de Investigaci\'on y Desarrollo (ANID) of Chile.
\bibliographystyle{elsarticle-num} 

\end{document}